\documentclass{elsart}
\usepackage{epsfig,graphics,times}
\usepackage{picinpar}
\usepackage{wrapfig}
\usepackage{floatflt}
\usepackage{enumerate}
\usepackage{color}

\begin{document}
\newcommand{\pbar}{$\overline{\mbox{p}}$}
\newcommand{\ds}[1]{\displaystyle{#1}} 
\renewcommand{\labelitemi}{$\triangleright$} 
\begin{frontmatter}
\title{Spatial Resolution\\ of Double--Sided Silicon Microstrip Detectors\\ for the PAMELA Apparatus}
\large{The PAMELA tracker collaboration:}
\author[Firenze]{S. Straulino\corauthref{samuele}}\ead{straulino@fi.infn.it}
\author[Firenze]{O. Adriani} 
\author[Firenze]{L. Bonechi}
\author[Firenze]{M. Bongi}
\author[Firenze]{S. Bottai}
\author[ifac]{G. Castellini} 
\author[Firenze]{D. Fedele}
\author[Firenze]{M. Grandi}   
\author[Firenze]{P. Papini} 
\author[Firenze]{S. B. Ricciarini} 
\author[Firenze]{P. Spillantini} 
\author[Firenze]{F. Taccetti}
\author[Firenze]{E. Taddei}
\author[Firenze]{E. Vannuccini}
\vspace*{-0.1cm}
\address[Firenze]{Universit\`a di Firenze and INFN Sezione di Firenze, via Sansone, 1 -- 50019 Sesto Fiorentino (Italy)}
\address[ifac]{Istituto di Fisica Applicata N. Carrara, CNR, via Panciatichi, 64 -- 50127 Firenze (Italy)}
\corauth[samuele]{Corresponding author.}

\begin{abstract}
The PAMELA apparatus has been assembled and it is ready to be launched in a satellite mission to study mainly the antiparticle component of cosmic rays. In this paper the performances obtained for the silicon microstrip detectors used in the magnetic spectrometer are presented. This subdetector reconstructs the curvature of a charged particle in the magnetic field produced by a permanent magnet and consequently determines momentum and charge sign, thanks to a very good accuracy in the position measurements (better than $3 \, \mu$m in the bending coordinate). A complete simulation of the silicon microstrip detectors has been developed in order to investigate in great detail the sensor's characteristics. Simulated events have been then compared with data gathered from minimum ionizing particle (MIP) beams during the last years in order to tune free parameters of the simulation. Finally some either widely used or original position finding algorithms, designed for such kind of detectors, have been applied to events with different incidence angles. As a result of the analysis, a method of impact point reconstruction can be chosen, depending on both the particle's incidence angle and the cluster multiplicity, so as to maximize the capability of the spectrometer in antiparticle tagging. 
\end{abstract}
\end{frontmatter}


\section{Introduction} 

\begin{figure}
\hspace*{-0.8cm}
 \begin{minipage}{7.5cm}
  \begin{center}
   \includegraphics[width=7.7cm]{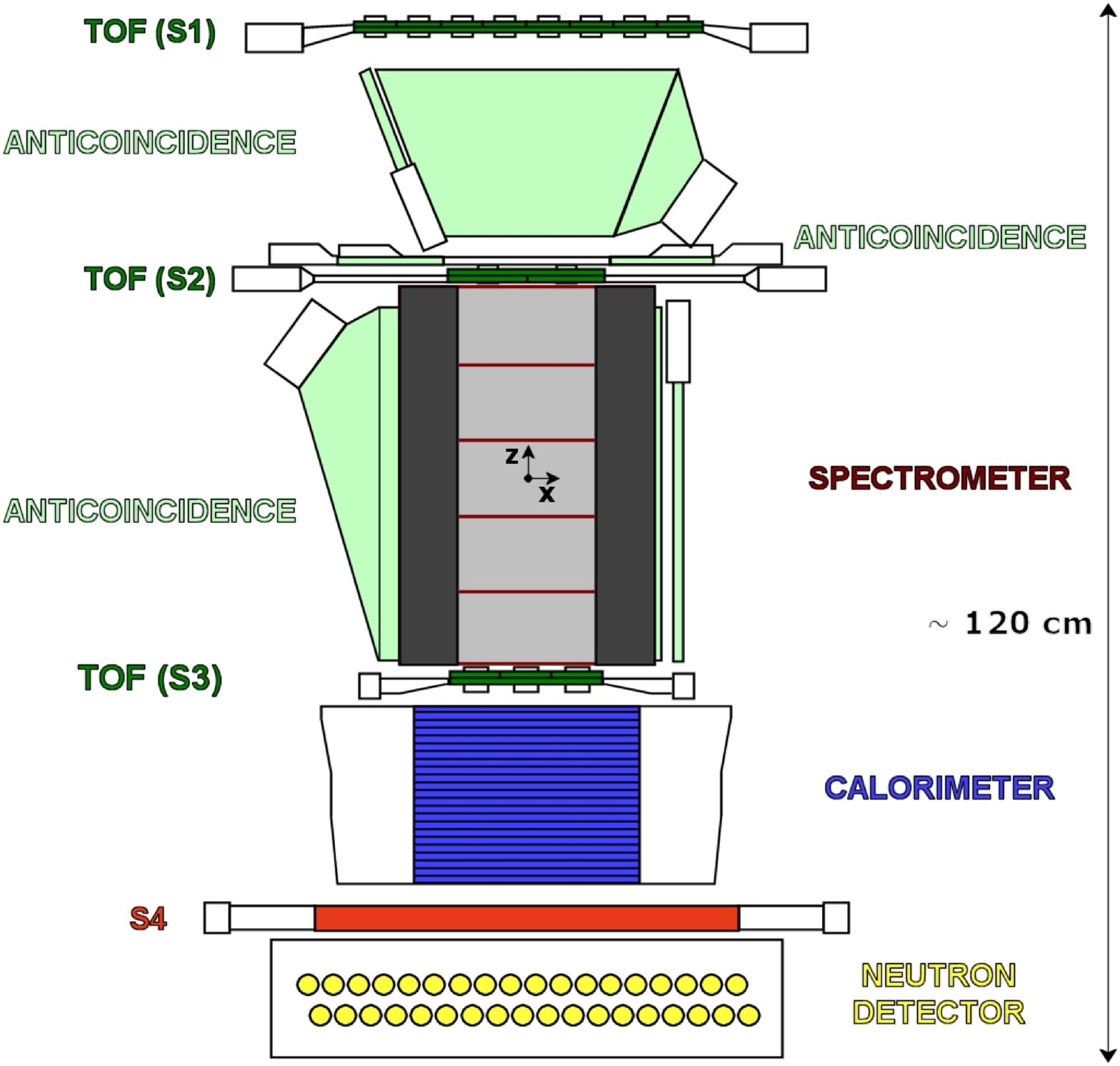}
  \end{center}
 \end{minipage}
\hspace*{-0.1cm}
 \begin{minipage}{7.5cm}
  \begin{center}
   \includegraphics[width=7.3cm]{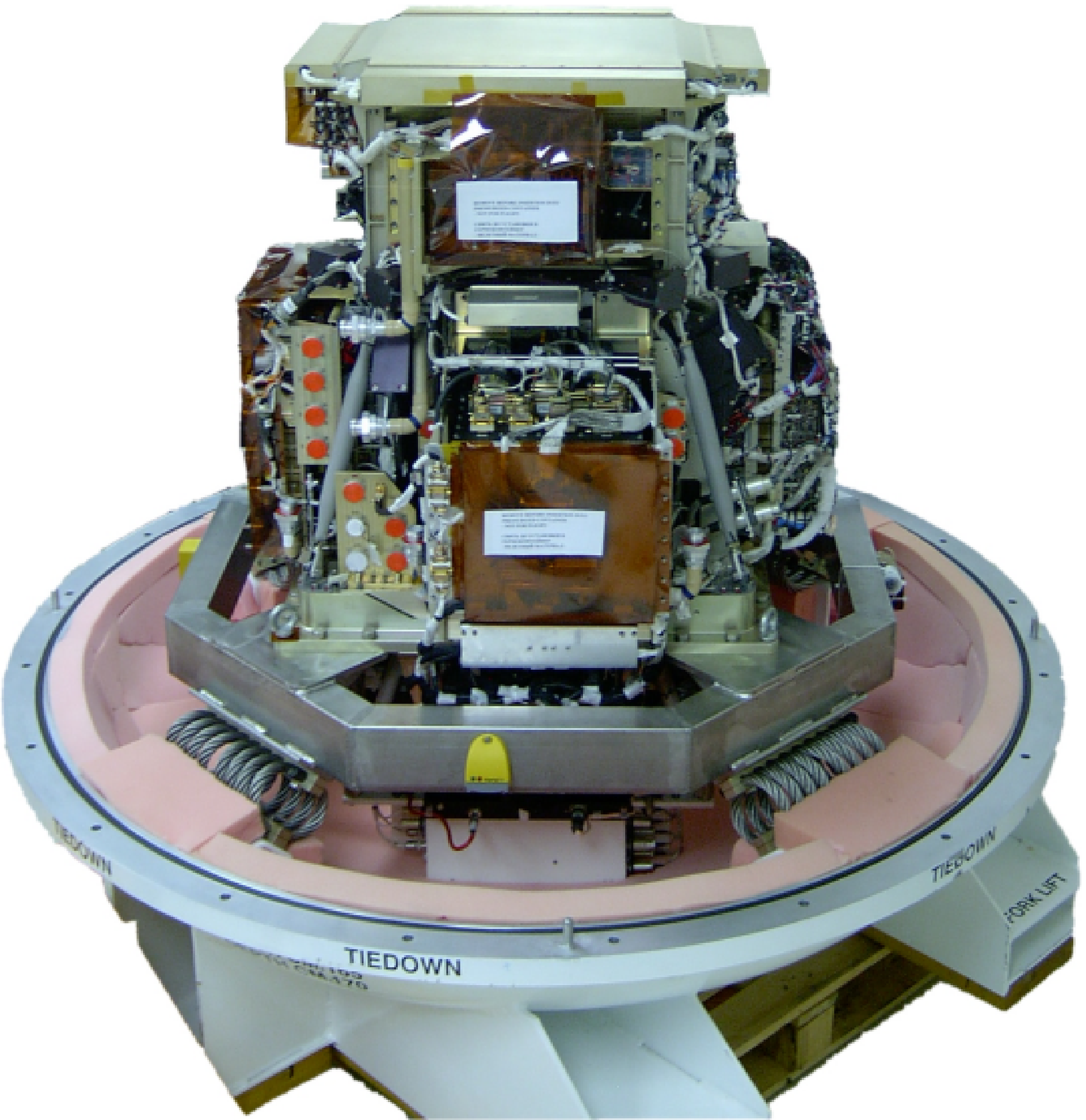}
  \end{center}
 \end{minipage}
 \begin{center}
  \begin{minipage}{13cm}
   \caption{Schematic lateral view of the PAMELA detector ({\em left}) and a photograph of it ({\em right}) taken before the delivery of the instrument to the Russian space company. The geometrical acceptance of the detector is 20.5~$\mbox{cm}^2\mbox{sr}$.}
   \label{fig:pamela}
  \end{minipage}
 \end{center}
\end{figure}

The PAMELA telescope~\cite{npb113,nima478} will be put in orbit within the 2005 on board of the Resurs DK1 Russian satellite for a three--year--long mission on a \mbox{quasi--polar} orbit ($70.4$~deg. inclination, $350$ to $600$~km height) to study the cosmic ray flux, with a special interest on the antimatter component. The detector is composed of several subsystems, schematically shown in fig.~\ref{fig:pamela}: a Time of Flight (ToF) apparatus, which also provides the trigger signal, a solid state Magnetic Spectrometer~\cite{nima485,bonechi}, surrounded by an anticoincidence shield, and an Electromagnetic Calorimeter~\cite{nima487} in which single--sided silicon detector planes are interleaved with tungsten absorber up to a total thickness of about $16$ radiation lengths. Anticoincidence scintillators define the external geometry of the detector and their signals will be exploited  in the off--line rejection of spurious tracks; below the calorimeter another scintillator plane (S4) and a Neutron Detector can provide additional information when showers are not fully contained in the calorimeter.  

The whole detector can discriminate protons and electrons from their antiparticles and provide energy measurements; also $Z>1$ nuclei may cross the entire spectrometer and consequently can be detected. Antimatter acquisitions will cover the energy range $80$~MeV--$190$~GeV for antiprotons and $50$~MeV--$270$~GeV for positrons, so extending the current upper limit (about $50$~GeV) given by balloon--borne missions~\cite{apj561}. In addition, the long duration of the flight will provide larger statistics in particular in the high--energy range, where the power--law spectrum of cosmic rays requires either large acceptances or long exposure times\footnote{The differential flux of galactic cosmic rays at $1$ Astronomical Unit from the Sun is proportional to $E^{-2.7}$ in the energy range $E \simeq 10 - 10^{6} \, \mbox{GeV/n}$~\cite{simpson:1983}.}. The obtained high--energy antiparticle flux will constrain the models describing the origin of antimatter in the Universe. The apparatus can be exploited also to find dark--matter signatures through the detection of high--energy antiprotons originating from neutralino annihilations~\cite{boeziopearce}.

\section{The magnetic spectrometer} 
\subsection{Magnet}

The magnetic spectrometer is the core of the PAMELA apparatus: it is based on a permanent magnet and consists of six detector planes which measure both the impact coordinates of the incoming particles. The magnet is made of a Nd--Fe--B alloy, with a high value of the residual magnetization (about $1.3$~T). Blocks of magnetic material define a rectangular cavity ($132 \times 162 \, \mbox{mm}^{2}$) where particles are detected. Here the magnetic field is roughly uniform and oriented along the Y coordinate in the PAMELA reference frame (fig.~\ref{fig:pamela}). As a consequence, particles are bent in the XZ plane within the cavity, due to the Lorentz force ${\bf F} = q{\bf v} \times {\bf B}$. Five identical magnetic modules, each one $80$~mm high, are superimposed each other and interleaved with six detector planes, which are inserted in the magnetic tower by means of dedicated slits. The magnetic field in the centre of the cavity is $0.48$~T. Measurements of the three components of the magnetic field have been taken at a fixed pitch in about $70 \,000$ points inside the cavity. Such values will be used during the off--line analysis to precisely determine the particle's momentum through a numerical integration of its equation of motion inside the cavity.

\subsection{Silicon detectors}

When the characteristics of the PAMELA experiment were studied, the main requirements of the detectors to be used in the magnetic spectrometer were defined. Essentially they can be listed as in the following:

\begin{enumerate}[{\em a)}]

\item provide two coordinates per detector;
\item maximize the spatial resolution, in particular for the bending coordinate;
\item minimize the multiple scattering.

\end{enumerate}
 
\begin{figure}
 \begin{center}
  \begin{minipage}{11cm}
  \begin{center}
  \frame{\includegraphics[width=10cm]{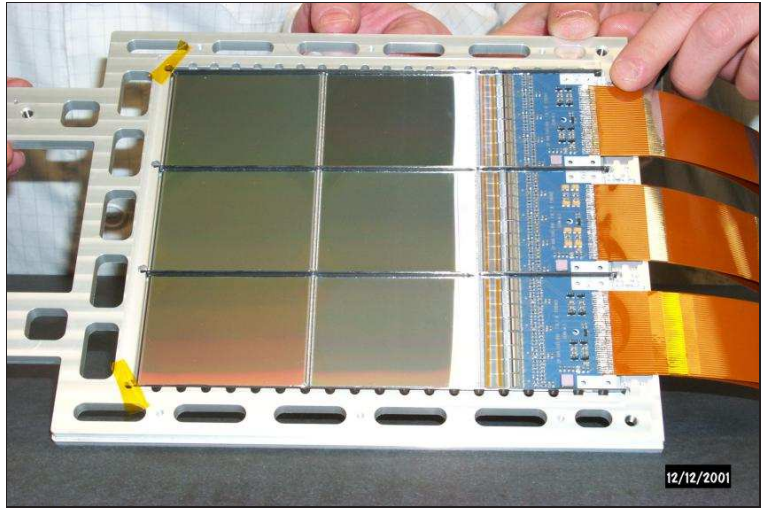}}
  \caption{Photograph of a silicon detector plane.}
  \label{fig:photoplane}
  \end{center}
  \end{minipage}
 \end{center}
\end{figure}

The best candidates to comply with such requirements appeared to be double--sided silicon microstrip detectors with a small strip pitch, associated with low--noise electronics. The multiple scattering was limited by excluding any dead layer above or beneath the detectors. In the resulting configuration six detectors, manufactured by Hamamatsu Photonics~\cite{hamamatsu}, are arranged in each plane (fig.~\ref{fig:photoplane}). A detector is made of a $300 \,\mu$m thick $n$--type silicon wafer, with $p$--type strips implanted at $25.5 \, \mu$m pitch on the junction side. These strips are used to determine the X coordinate of the incoming particle\footnote{The X coordinate is related to the particle's curvature: for this reason the X view of the detector is also called the \em{bending view}.}. $n^{+}$--type strips, orthogonal to the previous ones, are implanted with $66.5 \, \mu$m pitch on the Y (ohmic) side. The read--out pitch on the junction side is actually $51 \, \mu$m, since a $p$ strip out of two is left floating: charge collected by that strip influences the output of the adjacent read--out channels~\cite{turchetta:1993}. On the ohmic side a $p$ blocking strip is added between two consecutive $n$ strips in order to increase the interstrip resistance. Integrated decoupling capacitors are present on both sides: they are obtained by depositing an insulating layer ($0.1 \, \mu$m $\mbox{SiO}_{2}$) and metallic strips above the implants. On the ohmic view a second metal layer, separated by another $\mbox{SiO}_{2}$ deposition, $5 \, \mu$m thick, brings the read--out metallic strips parallel to the junction side's ones. Both surfaces of the detectors are passivated; electrical signals can be taken out from metallic pads located at both ends of the strips. The depletion voltage given by the manufacturer is $60 \pm 20$~V and $80$~V bias will be supplied during the operational phase. The detectors are arranged in three modules named ladders: two sensors and a double--sided hybrid circuit form each ladder. These elements are glued in such a way that X side strips meet at their ends: electrical connections among the corresponding strips are provided by $17 \, \mu$m wire bonds on both sides. On the ohmic view two distinct strips, about $70 \, \mbox{mm}$ apart, are connected to the same electronic channel, so giving rise to a position ambiguity, that can be easily resolved on the basis of the other PAMELA's subdetectors. The percentage of bad strips, due mainly to unconnected or noisy channels, ranges between $1$\% and $5$\% for all the ladders. The mechanical structure of the plane is shown in fig.~\ref{fig:photoplane} and it minimizes the multiple scattering since no additional dead layer is present on the particle's trajectory. Nevertheless, in order to increase the rigidity of the structure, four carbon--fibre bars (visible in the photograph) with very high Young modulus (about $300$~GPa) are glued at both sides of each ladder. The resulting plane has the first resonance frequency located beyond $300$~Hz and it has survived vibrational tests whose intensities were twice those expected during the launch of the satellite~\cite{nima511}. A summary of geometrical and electrical characteristics of the detectors is given in table~\ref{tab:charact}.

\begin{table}
\begin{center}
\begin{tabular}{|l|c|c|}
\multicolumn{1}{c}{}& \multicolumn{1}{c}{Junction side} & \multicolumn{1}{c}{Ohmic side} \\ 
\hline
Strip pitch and type     & $25.5 \, \mu$m, p & $66.5 \, \mu$m, n \\
\hline
Read--out pitch          & $51 \, \mu$m   & $66.5 \, \mu$m \\
\hline
Integrated capacitors    & yes & yes \\
\hline
Double metal             & --  & yes \\ 
\hline
p--stop strip            & --    & one \\
\hline
Bias resistor            & punch--through, $>50 \, \mbox{M}\Omega$ &
polysilicon, $>10 \, \mbox{M}\Omega$ \\
\hline
Total capacitance to GND & $\le 10$~pF & $\le 20$~pF\\
\hline
Leakage current          & \multicolumn{2}{|c|}{$<1 \, \mu$A per sensor} \\
\hline
Depletion voltage        & \multicolumn{2}{|c|}{$60 \pm 20$~V} \\
\hline
\end{tabular}
\end{center}
\vspace*{0.3cm}
\caption{Geometrical and electrical characteristics of the silicon microstrip detectors.}
\label{tab:charact}
\end{table}

\subsection{Front--end electronics}
The front--end electronics is housed on the double--sided hybrid: it is based upon the VA1~\cite{va1} chip, which consists of $128$ charge preamplifiers, shapers and sample--and--hold stages in series with a multiplexer controlled by a shift register. The dynamic range of the chip corresponds approximately to $\pm 10$~MIP's. The analog signal on each channel is then digitized by means of a $12$--bit ADC per ladder view. As known in literature~\cite{spieler:1985}, the noise observed in a detector--preamplifier chain is essentially related to four parameters, depending on both the sensitive element and the associated front--end electronics: the bias resistance $R_{B}$, the leakage current $i_{L}$, the total capacitance $C$ to ground seen by the preamplifier and the shaping time $\tau$. For a given shaping time ($\tau = 1 \, \mu$s in our case), the noise is minimized by a large value of the bias resistance and by small values of leakage current and load capacitance. Let us briefly discuss each contribution as it appears in our configuration. Bias resistors are used to connect each strip to the guard rings, which are kept at a fixed potential. On the junction side the punch--through technology~\cite{ellison:1989} with an additional foxfet electrode~\cite{allport:1991} has been exploited to have $R_{B} > 50 \, \mbox{M}\Omega$; on the ohmic side polysilicon resistors have been used and the corresponding resistance is greater than $10 \,\mbox{M}\Omega$. The measured leakage current is of the order of $1 \, \mbox{nA/strip}$. The $R_{B}$ value and the limit on $i_{L}$ have been designed to give a negligible noise, compared to the contribution from the capacitive load seen by the preamplifier. The requirement on this parameter imposed to the detector's manufacturer was $C \le 10 \, \mbox{pF} \, (20\, \mbox{pF})$ for the junction (ohmic) side of a sensor. 

The VA1 Equivalent Noise Charge, according to the manufacturer's specifications, can be written in the nominal working conditions as: 

\begin{equation}
\mbox{ENC}\;=\; a + b \cdot C \;=\; 180 \, \mbox{e}^{-} + 7.5 \, \mbox{e}^{-} \! \cdot C \; \mbox{(pF)}
\label{eq:enc}
\end{equation}

Since a minimum ionizing particle creates on the average $27500 \, \mbox{e}^{-}$ in a $300 \, \mu$m thick silicon layer, from eq.~\ref{eq:enc} we expect for the signal--to--noise ratio $S/N$ on both sides values of this order:
$ \langle S/N \rangle_{X} \;  \simeq  \; 83 \, ; \;\;
  \langle S/N \rangle_{Y} \;  \simeq  \; 57 \, . $
Actually from the acquisition taken on beam test data we have obtained (fig.~\ref{fig:snxy}):
$ \langle S/N \rangle_{X} \;  \simeq  \; 54 \, ; \;\;
  \langle S/N \rangle_{Y} \;  \simeq  \; 23  \, .$
Therefore the VA1 chip in the real working condition shows a lower performance: this can be explained remembering that we are operating the preamplifier with reduced power consumption ($1$~mW per channel instead of $1.3$~mW per channel) by properly changing its bias conditions. We were forced to operate in this way because of the constraints imposed by the satellite. Nevertheless the signal--to--noise ratios obtained with lower power consumption are good enough to comply with the requirements of spatial resolution expected for the apparatus. We did also a check of the change of performance for the different VA1's working condition by measuring the noise obtained from the chip without any load attached (i.e. the $a$ parameter in eq.~\ref{eq:enc} has been determined): it resulted $232 \, \mbox{e}^{-}$ instead of the nominal value $180 \, \mbox{e}^{-}$.

\section{Reduction and analysis of tracking system data} 
\label{sec:reduction}

$36 \,864$ electronic channels are present in the whole tracker. If an average estimated trigger rate of $12$~Hz~\cite{nima511} is assumed for in--orbit acquisitions (including spurious events, such as those related to interactions in external dead materials) a total amount of more than $50$~GB/day can be inferred only for the spectrometer. This value is larger than the maximum available bandwidth for data transmission to the ground stations ($10$~GB/day). A compression procedure is then applied on--line to the raw data: it consists of a Zero Order Predictor--like algorithm, completed by a {\it cluster finder} which preserves the set of signals on adjacent channels corresponding to the passage of an ionizing particle. The amount of data to be transmitted is then reduced to about $5$\% of the initial size without degrading the spatial resolution of the instrument. In the off--line analysis values of non--transmitted channels are reconstructed on the basis of the transmitted ones. 

Data acquired on satellite consist of {\em calibration runs}, periodically repeated, and {\em physics runs}. In the first ones (acquisitions without particles) the pedestal ($PED$) and the intrinsic noise ($N$) of each channel are evaluated. In the physics runs the pedestal values are used to obtain, by a recursive procedure, the common noise ($CN$) of each VA1 chip, different from an event to another. The true signal $S$ of a given channel is then obtained from the $12$--bit ADC value ($ADC$) as:

\[
S = ADC - PED - CN
\]
When all the signals of all the electronic channels are available, those corresponding to the passage of an ionizing particle are selected through the value of the variable:

\begin{figure}
 \begin{center}
  \begin{minipage}{12.5cm}
  \begin{center}
  \includegraphics[width=12cm, bb=0 300 596 681, clip]{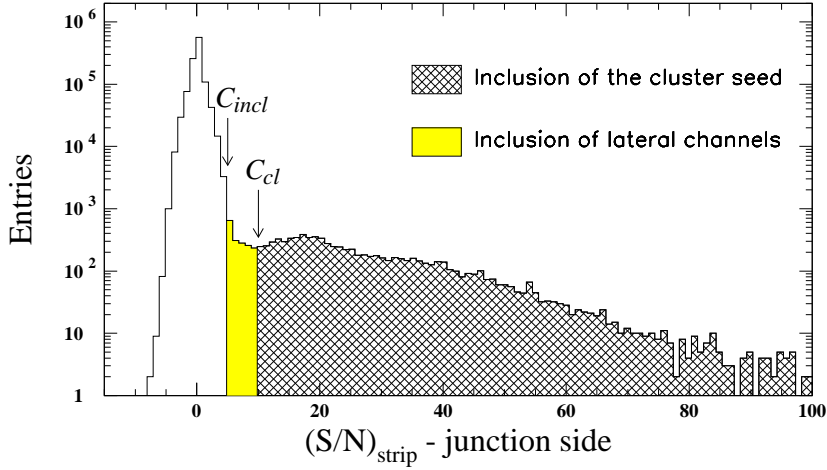}
  \caption{Distribution of the $R = (S/N)_{strip}$ variable for all the read--out strips on the junction side of a ladder, as obtained from MIP tracks, orthogonal to the detector, on a test beam. The peak centred in zero is the noise distribution, while particle signals are contained in the high--$S/N$ tail. The threshold $C_{cl} = 10$ selects the strip whose signal is expected to be the maximum of the cluster, while $C_{incl} = 5$ determines whether lateral strips have to be included in it.}
  \label{fig:cut_x}
  \end{center}
  \end{minipage}
 \end{center}
\end{figure}

\begin{figure}
 \begin{center}
  \begin{minipage}{12.5cm}
  \begin{center}
  \includegraphics[width=12cm, bb=0 0 596 395, clip]{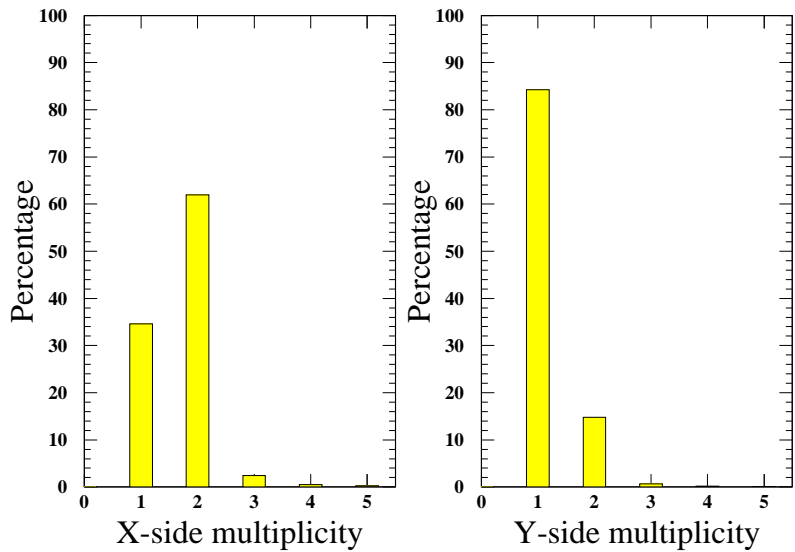}
  \caption{Percentage of events with given multiplicity on both sides of the sensors (orthogonal MIP tracks from a test beam).}
  \label{fig:multxy}
  \end{center}
  \end{minipage}
 \end{center}
\end{figure}

\begin{figure}
 \begin{center}
  \begin{minipage}{12.5cm}
  \begin{center}
  \includegraphics[width=12cm, bb=0 310 596 655, clip]{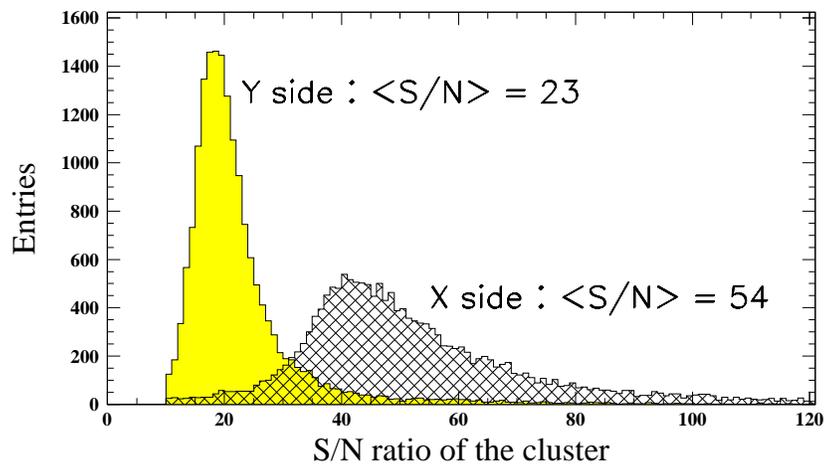}
  \caption{Distribution of the signal--to--noise ratio of the cluster for both X and Y sides of a ladder (orthogonal MIP tracks from a test beam).}
  \label{fig:snxy}
  \end{center}
  \end{minipage}
 \end{center}
\end{figure}

\[
R = S / N
\]
The $R$ variable represents the signal--to--noise ratio of a given strip for the considered event. A $R$--value greater than a chosen threshold $C_{cl}$ reveals the so--called {\it cluster seed}; channels adjacent to the maximum on both sides are then included in the cluster as far as their signals fall below a second, lower threshold $C_{incl}$. When a channel has a signal below the inclusion level $C_{incl}$, search for further strips stops on that side of the cluster with respect to the maximum. The values $C_{cl} = 10$ and $C_{incl} = 5$ have been used for this analysis; their positions in the distribution of the $R$ variable are shown in fig.~\ref{fig:cut_x} for the junction side of a typical detector. The number of strips in a cluster, according to the above inclusion rule, is called its {\em multiplicity}; the distribution of this variable on both sides is given in fig.~\ref{fig:multxy} for tracks orthogonal to the sensors from a test beam. Following ref.~\cite{turchetta:1993}, the {\em signal--to--noise ratio of the cluster} is defined as the sum of the $S/N$ ratios of all its strips: the corresponding distribution is given in fig.~\ref{fig:snxy} for the same set of events as in fig.~\ref{fig:multxy}. 

\section{Detector simulation} 

\begin{figure}
 \begin{center}
  \begin{minipage}{12.5cm}
  \begin{center}
  \includegraphics[width=11cm]{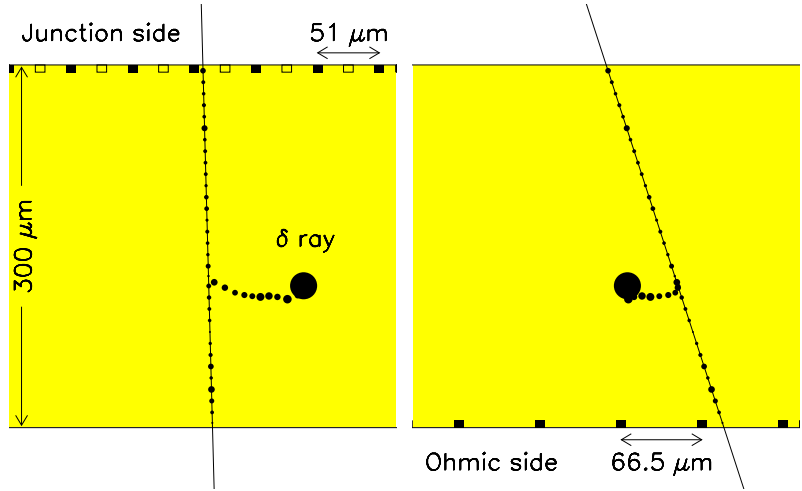}
  \caption{Simulated track by a MIP in the silicon detector with emission of a
  $\delta$ ray. The event is seen on two planes orthogonal to the strips
  of junction ({\em left}) and ohmic ({\em right}) sides.
  The black rectangles represent the read--out strips;
  the blank ones are the floating strips of the X, junction view.}
  \label{fig:evplot_o}
  \end{center}
  \end{minipage}
 \end{center}
\end{figure}

The developed simulation includes the complete chain of physical processes produced in consequence of a crossing particle, from ionization of silicon to the digitized output signals. The first step is achieved through GEANT~\cite{geant}, which reproduces the charge generation along the track of the ionizing particle. In fig.~\ref{fig:evplot_o} a simulated track originating from a MIP in the silicon detector is shown for two orthogonal cross sections: each point along the track represents a charge packet and its area is proportional to the energy loss. Tracking is carried out using finite steps, whose maximum length has been set to be $10 \, \mu$m (corresponding to at least $30$ charge packets per track inside the $300 \, \mu$m thick detector) to exploit a fine granularity in the energy deposition.

In a second phase the charge packets created in the sensitive volume are collected on the strips: their behaviour can be described as due to both drift and diffusion. The drift velocity $\bf{v}$ of charge carriers is given by the product of the mobility $\mu$ and the electric field $\bf{E}$: $\,\bf{v} = \mu \cdot \bf{E}\;$. Moreover the charge carriers diffuse in the detector material: the point--like charge packets widen according to a Gaussian distribution. As a consequence, the standard deviation of the packet's spatial distribution at time $t$ after its generation can be written as 
\begin{equation}
\sigma = \sqrt{2Dt}
\label{eq:diffus}
\end{equation}
$D$ being the diffusion coefficient, which depends mainly on temperature and characteristics of the material~\cite{samueletesi}. All the charge packets, located along the track, are collected on the strips and their spatial distributions, that have different widths, contribute to the creation of a cumulative packet on the surface of the detector. To give an idea of the transversal size of the charge cloud, in our configuration the standard deviation of a packet that has covered half detector thickness is about $5 \, \mu$m and it gets the same value for electrons and holes~\cite{peisert:1992}. 

The simulation of the drift process is based on the knowledge of the electric field $\bf{E}$ inside the detector. Along a coordinate ($z$ in our case) orthogonal to the {\em p--n} junction $\bf{E}$ can be simply obtained from the solution of the $1$--dimensional Laplace's equation, as done in ref.~\cite{belau:1983}. In case of overdepleted detectors (i.e. if the bias voltage $V_{bias}$ is greater than the depletion voltage $V_{depl}$) the electric field intensity can be written at a distance $z$ from the junction as:

\[ E(z) = -2 \frac{V_{depl}}{d^{\,2}} (d-z) - \frac{V_{bias}-V_{depl}}{d} \]

\noindent
where $d$ is the thickness of the detector, corresponding to the depletion zone.

\begin{figure}
 \begin{center}
  \begin{minipage}{12.5cm}
  \begin{center}
  \includegraphics[width=12.5cm,bb=0 0 596 395,clip]{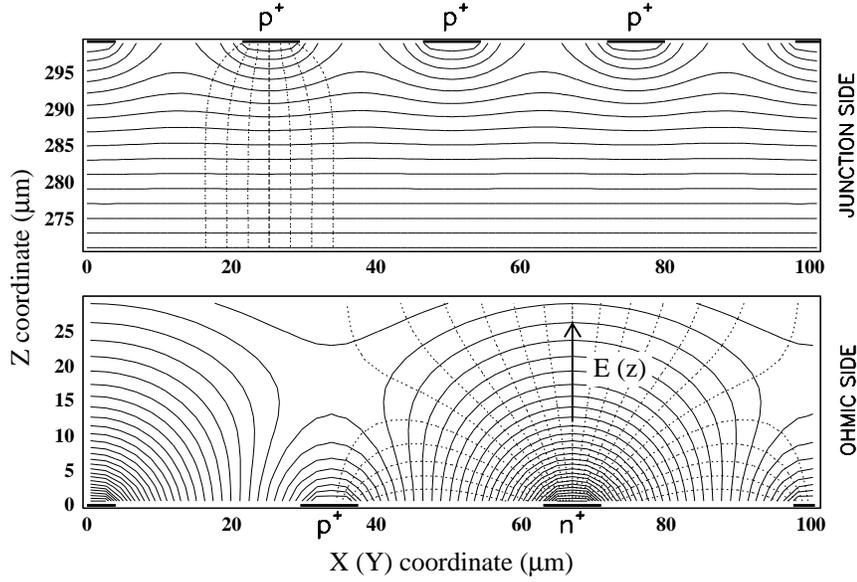}
  \caption{Behaviour of equipotential lines (solid) and electric field 
  lines (dashed) in proximity of junction ({\em top}) and ohmic ({\em bottom}) sides of the
  sensor.}
  \label{fig:laplace}
  \end{center}
  \end{minipage}
 \end{center}
\end{figure}

\begin{figure}
 \begin{center}
  \begin{minipage}{12.5cm}
  \begin{center}
  \frame{\includegraphics[width=10cm, bb=50 35 550 370, clip]{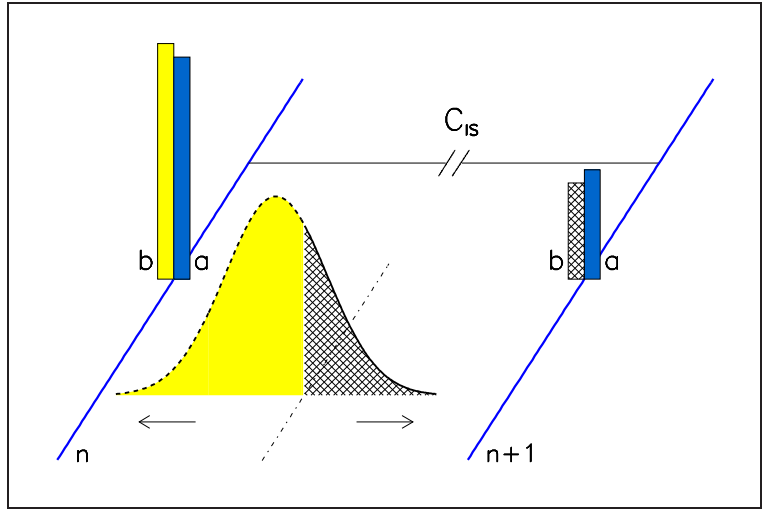}}
  \caption{Mechanism of division of a Gaussian charge packet between two subsequent strips: each one collects a charge signal corresponding to the fraction of area closer to itself. The presence of the interstrip capacitance $C_{IS}$ gives rise to a rearrangement of the charge: a fraction of each signal appears on the output of the adjacent channels. As an example signals before ($b$) and after ($a$) the introduction of this capacitive coupling are depicted in figure.}
  \label{fig:packet}
  \end{center}
  \end{minipage}
 \end{center}
\end{figure}

In the real case, the $3$--dimensional map of the electric field is in principle necessary to determine the path in silicon of the charge carriers. Thanks to the symmetry of the system, this map has been obtained by solving by a numerical method the $2$--dim. Laplace's equation for each side of the sensor~\cite{samueletesi}. In fig.~\ref{fig:laplace} the equipotential lines and the electric field lines are shown for regions close to X and Y sides of the sensor, corresponding to $1/10$ of the full detector thickness. For the junction side (top) the  $\bf{E}$--lines are orthogonal to the surface of the sensor, except in a thin layer ($\sim 10~\mu$m) close to the junction. Since the electric field lines are essentially straight lines except in a small zone close to the collection plane, the drift of each charge packet is reproduced in the simulation simply as a ``translation'' of it towards the strips and the diffusion is considered as a Gaussian enlargement of the packet up to the final width given by eq.~\ref{eq:diffus}, where $t$ is now the collection time of the packet. Finally, this charge packet is divided between adjacent strips, as illustrated in fig.~\ref{fig:packet}. 

On the ohmic side (bottom) the  calculation has been performed considering the effect of the $p$ blocking strip. According to measurements reported in literature, when the junction is reverse biased such structures reach at equilibrium a potential which mainly depends on both the strip geometry and the doping level. Values $10$--$15$~V lower than those of the adjacent $n^{+}$ strips have been reported~\cite{matheson:1995,richter:1996}. If the $p$--stop potential is introduced as boundary condition for the solution of the Laplace's equation, the field lines assume a shape similar to that shown in the second graph of fig.~\ref{fig:laplace}: around the $p$--stop strips, a zone with a lower density of field lines can be observed. When particles cross the sensor in this region, charge carriers spend more time in the material before being collected on the $n^{+}$ strips: their diffusion widths increase consequently. In order to reproduce this effect in the simulation, the width of the resulting diffusion cloud has been parameterized for the ohmic side as a function of the interstrip position.

Capacitive couplings between adjacent strips are also taken into account: a fraction of the signal appearing on each channel is actually visible on the adjacent output channels (fig.~\ref{fig:packet}). In a first--order approximation, this fraction corresponds to the ratio $r=C_{IS}/C_{DEC}$ between the first--neighbour interstrip and the decoupling capacitances~\cite{turchetta:1993}. A second--neighbour coupling has also been introduced on the junction side for a fine tuning with data. Values of the fraction of transferred charge for both junction and ohmic sides have been obtained by tuning simulation and data. On the junction side an additional difficulty is the presence of the intermediate, floating strip: charge collected by it is partly lost to the ohmic side, while the surviving fraction is shared upon the adjacent strips. Also this effect has been tuned on data.

The electronic noise is the last physical effect to be reproduced in the simulation. To this purpose Gaussian fluctuations have been superimposed to the signals arising in the simulated electronic channels: the corresponding standard deviations have been extracted from a data file acquired on a test beam.

\section{Effects of a magnetic field}

The presence of a magnetic field in the region where detectors operate influences the collection of the charge on the strips. Assume to have a uniform magnetic field parallel to the strip direction and a particle beam orthogonal to the sensor. Due to the Lorentz force on the charge carriers, electrons and holes produced along the track of a crossing particle are collected at an angle $\vartheta_{L}$ (Lorentz angle) with respect to the track, which in the International System can be evaluated as:

\[ \vartheta_{L} = \mu_{H} B \]

\noindent
where $B$ is the value of the magnetic induction and $\mu_{H}$ is the so--called Hall mobility~\cite{peisert:1992}. In our configuration the Lorentz angle is about $1$~deg. if the value $\mu_{H} = 0.0310 \,\mbox{m}^{2} / \mbox{Vs}$~\cite{peisert:1992} is used along with a mean value of the magnetic field of $0.45$~T. The vector $\bf{B}$ is orthogonal to the Y strips in PAMELA, giving rise to a Lorentz angle $\vartheta_{L}$ in the XZ plane (and therefore affecting the measured position on the junction side only). In this plane an effective angle $\vartheta_{eff} = \vartheta + \vartheta_{L}$ should be considered, instead of the incidence angle $\vartheta$. The magnetic field has not been considered in the simulation because our main present interest is the study of the intrinsic detector performances. Nevertheless the magnetic effect can be easily introduced in the existing code, if the characteristics of the whole spectrometer, in presence of straight and inclined tracks, will be investigated. 

\section{Tuning of the simulation on data} 
\label{sec:tuning}

A comparison between simulation and data coming from a test beam with MIP tracks orthogonal to the detector ($200$~GeV protons gathered at CERN--SPS) enables us to check the correctness of our work. In particular the cluster charge distribution, the average shape of the cluster and the $\eta$ function for data and simulation have been used for a fine tuning of some parameters used in the simulation.

\begin{figure} 
 \begin{center}
  \begin{minipage}{12.5cm}
  \begin{center}
  \includegraphics[width=12cm, bb=0 0 596 395, clip]{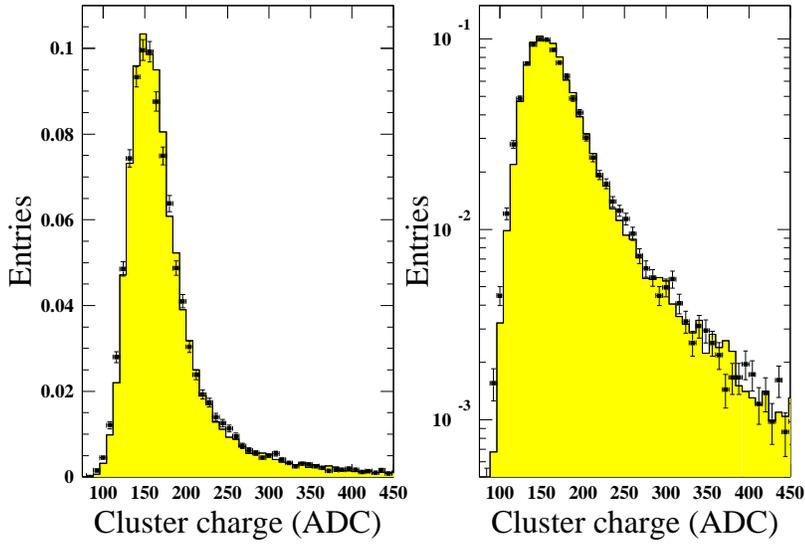}
  \caption{Cluster charge distribution (in ADC counts) on the junction side for data (points) and simulation (line). A good agreement is observed in both linear ({\em left}) and logarithmic ({\em right}) scales.}
  \label{fig:clucha}
  \end{center}
  \end{minipage}
 \end{center}
\end{figure}

\begin{figure}
 \begin{center}
  \begin{minipage}{12.5cm}
  \begin{center}
  \includegraphics[width=12cm, bb=0 0 596 395, clip]{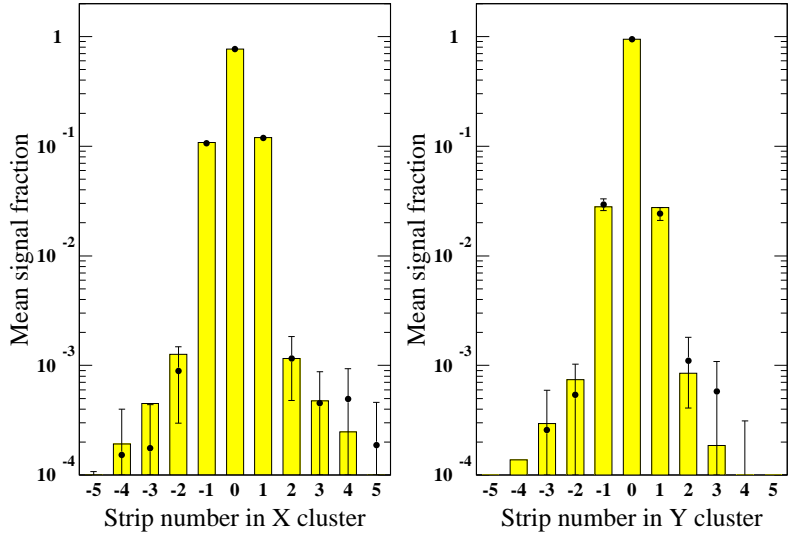}
  \caption{Comparison between the average shape of the cluster obtained from beam test data (points) and the corresponding simulated quantity (line) for both junction and ohmic sides (tracks orthogonal to the sensor). The data sample consists of about $20 \,000$ clusters.}
  \label{fig:clusha}
  \end{center}
  \end{minipage}
 \end{center}
\end{figure}

\begin{figure}
 \begin{center}
  \begin{minipage}{12.5cm}
  \begin{center}
  \includegraphics[width=12cm, bb=0 0 595 420, clip]{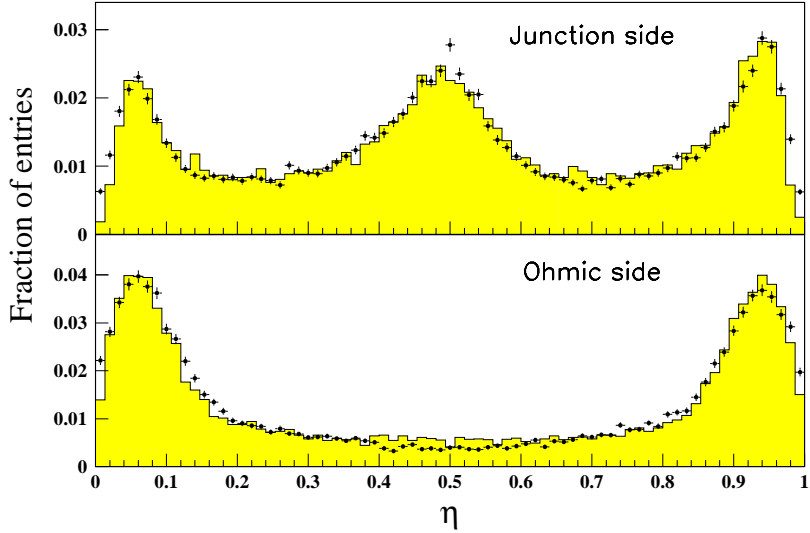}
  \caption{The $\eta$ distribution on both sides, for data (points) and simulation (line). In this data run tracks were approximately perpendicular to the surface of the detector; actually the asymmetry of the junction side distribution is due to a small inclination of the beam ($\sim 0.3$~deg.), reproduced in the simulation.}
  \label{fig:eta}
  \end{center}
  \end{minipage}
 \end{center}
\end{figure}

The distribution of the cluster charge, defined on the basis of the inclusion rules described in sect.~\ref{sec:reduction}, is reported in fig.~\ref{fig:clucha}. Provided charge is not lost outside the cluster, this quantity is proportional to the energy loss in the detector thickness. In fact, histograms in fig.~\ref{fig:clucha} follow a Landau distribution, as expected for MIP's crossing thin layers of material. The same agreement between data and simulation, shown in figure for the junction side, is observed on the ohmic view. The average shape of the cluster is shown in fig.~\ref{fig:clusha}: each graph has on the abscissa the strip number around the maximum of the cluster and on the ordinate the fraction of signal collected by the corresponding strip. All the clusters (about $20000$ in the data sample) are superimposed in the histogram to obtain an average behaviour. A left--right asymmetry (at $3 \sigma$ level) can be observed in figure on the junction side for the strips adjacent to the maximum: this difference could be explained as due to a small shift ($\sim 1 \, \mu$m) of the mask used to implant the strips in the silicon bulk with respect to the mask used to define the metallic strips. As a consequence, the capacitive coupling of a strip to the left neighbour is different from the coupling to the right one. This possibility has been confirmed by the manufacturer. The simulation shown in fig.~\ref{fig:clusha} includes such asymmetry. The good agreement between data and simulation represents a first significant test of the general model, since the average shape of the cluster is influenced by all the mechanisms of charge division and rearrangement on the strips (diffusion of charge carriers, capacitive couplings, signal--to--noise ratio).

The $\eta$ variable~\cite{turchetta:1993,belau:1983} is the basic parameter for our following analysis. If the pair of adjacent channels showing the highest signals are considered in each cluster, $\eta$ can be defined as:

\begin{equation}
\eta = \frac{R}{L+R}
\label{eq:etadef}
\end{equation}

\noindent
where $L$ and $R$ are respectively the signals of the left and right channel in the pair. The $\eta$ variable can be thought as an average (weighted by the signals $L$ and $R\,$) of the positions of two strips located in $0$ and $1$: if $\eta \sim 0$ or $\eta \sim 1$, it means that practically only one strip collects all the cluster signal. When this variable is reported on graph for all the clusters generated by a set of orthogonal particles uniformly distributed over the sensor, the characteristic behaviours shown in fig.~\ref{fig:eta} can be observed for junction and ohmic sides. This figure contains a comparison between the distributions of the $\eta$ variable for simulation (line) and beam test data (points). Peaks near $0$ and $1$ correspond to the presence of the read--out strips, while the peak near $0.5$ on the junction side corresponds to the intermediate floating strip: channels adjacent to such a strip collect approximately the same amount of charge (i.e. $L \simeq R$) when the floating strip has been hit by a particle. 

Connecting the peculiar shape of the $\eta$ distribution with the detector's characteristics is very instructive (see ref.~\cite{turchetta:1993}). The positions of the lateral peaks in the distributions are related to the capacitive coupling of a strip to the adjacent ones: larger the coupling, further the peaks are from $0$ and $1$, since the fraction of transferred charge becomes more and more considerable. The peak widths are related to the size of the diffused packet and to the $S/N$ ratio. Concerning the central peak on the X--side distribution, both the electronic noise and the loss of charge to the ohmic side (that resulted about $20$\% from a tuning with data) influence its width. 

Since the $\eta$--value is related to the main geometrical and electrical parameters of the detectors, it is a good probe to check if the code correctly reproduces the real detector response. A reasonable agreement between the $\eta$ distributions for data and simulation has been obtained by a fine--tuning of free parameters (fig.~\ref{fig:eta}). On the basis of this positive comparison, we are confident to use the simulated $\eta$ distribution in the procedure of impact point reconstruction, as explained in the following section.

\section{Position finding algorithms} 

A procedure that enables us to obtain the position of the incident particle on the sensor, once the signals induced over all the electronic channels are known, is called position finding algorithm. Such a method usually applies to clusters (identified on the basis of properly defined selection criteria) and allows to extract from them the particle impact point position. The general properties of algorithms for position measurements are well described in ref.~\cite{landi:2002,landi:2003}. A thorough description of the algorithms used for silicon microstrip detectors is given in ref.~\cite{turchetta:1993}; here we review only those our analysis is concerned with.  

\begin{itemize}

\item In the very simple {\it digital algorithm} the position of the strip which exhibits the biggest signal is assumed to be the incidence point of the particle.

\item Another intuitive algorithm is the {\it Centre Of Gravity (COG)}: the particle position can be estimated as the weighted average of the positions $x_{i}$ of the $m$ strips included in the cluster:

\begin{equation} 
x = \frac{\sum_{i=1}^{m} \, S_{i} x_{i}}{\sum_{i=1}^{m} \, S_{i}} 
\label{eq:ipr_cog}
\end{equation} 

\noindent
the weights $S_{i}$ being the corresponding signals.

\begin{figure}
\hspace*{-0.2cm}
 \begin{minipage}{7.5cm}
  \begin{center}
   \includegraphics[width=7cm, bb= 0 0 1534 2230, clip]{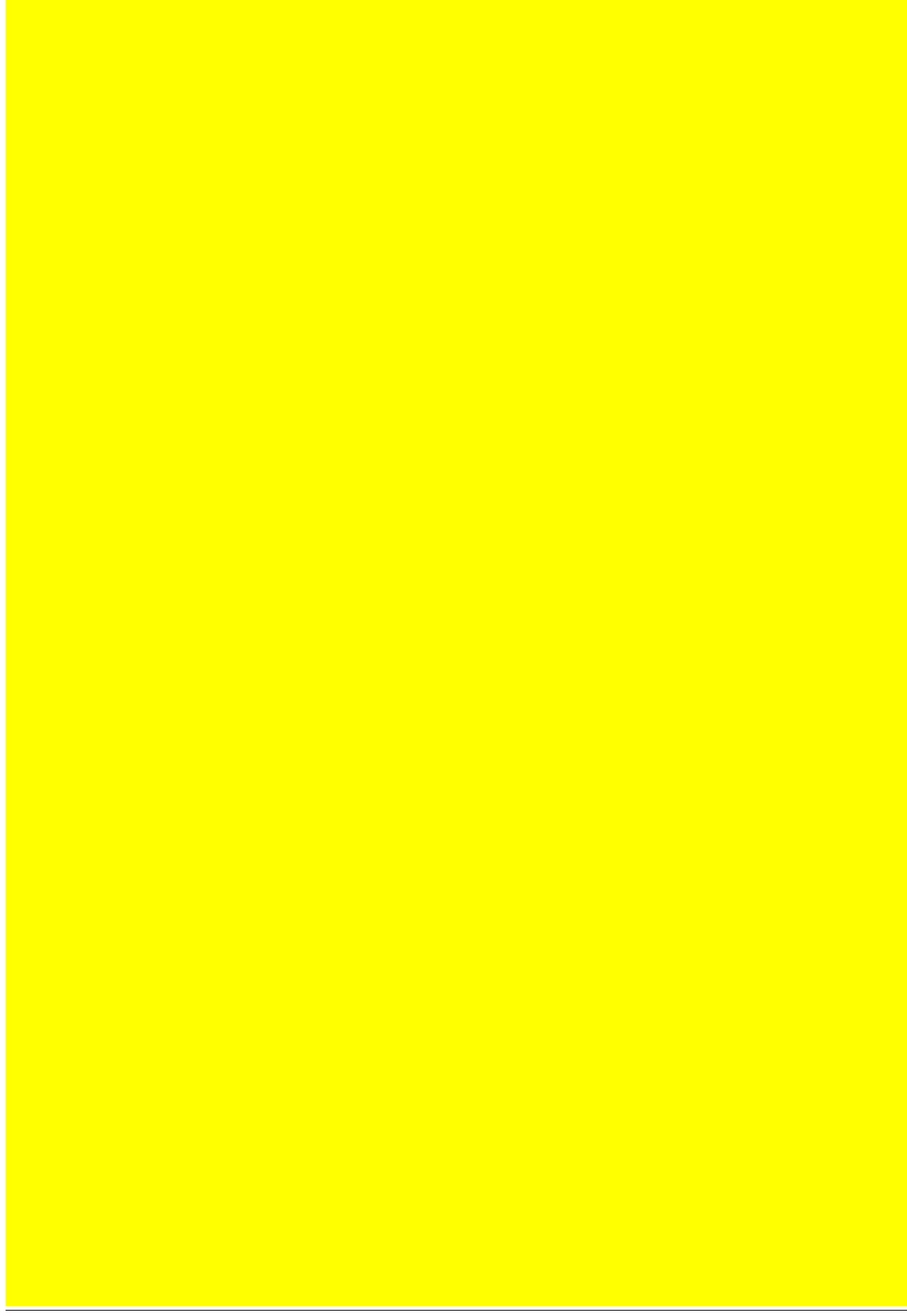}
  \end{center}
 \end{minipage}
\hspace*{-1.0cm}
 \begin{minipage}{7.5cm}
  \begin{center}
   \includegraphics[width=7cm, bb= 0 0 1534 2230, clip]{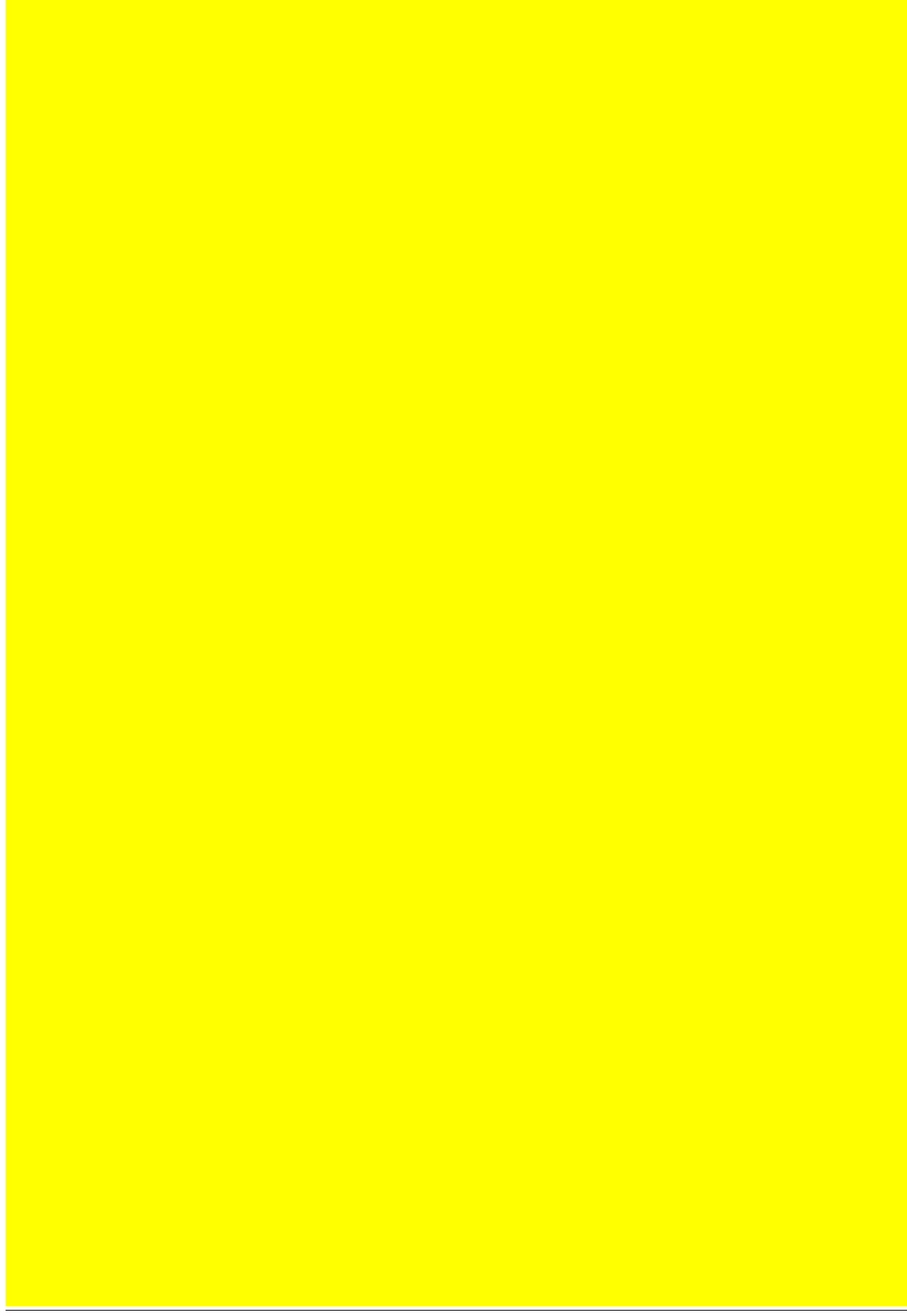}
  \end{center}
 \end{minipage}
 \begin{center}
\hspace*{+0.5cm}
  \begin{minipage}{12.5cm}
   \caption
{Simulation of two tracks inside the silicon detector and corresponding signals on the junction side. The typical clusters, obtained from tracks orthogonal to the detector, are shown. Channels are included in the clusters (grey bars) provided they satisfy the selection criteria described in sect.~\ref{sec:reduction}. On the left, the particle passes close to a read--out strip (black rectangle) which collects all the charge created in silicon: the corresponding cluster multiplicity is one. On the right, the particle crosses the detector near a floating strip (blank rectangle): the charge collected by it appears on the adjacent read--out channels and the corresponding cluster multiplicity is two.} 
   \label{fig:clusters}
  \end{minipage}
 \end{center}
\end{figure}

\item In case of orthogonal to the sensor or slightly inclined tracks one can think of limiting the previous sum to the pair of adjacent strips which collect the biggest signals. This is a reasonable approximation because, for small incidence angles, only one or two strips collect the whole charge generated in silicon by an ionizing particle. This is illustrated in fig.~\ref{fig:clusters}, where typical clusters for the junction side, corresponding to orthogonal tracks, are shown. Our detectors, on the X view, exhibit clusters whose multiplicity $m$ is one or two for about $98$\% of events in case of perpendicular tracks; higher multiplicity is associated with emission of energetic $\delta$ rays in silicon. If the events with $m = 1$ are separated from those with $m = 2$, the first ones are characterized by a worse accuracy in the reconstructed position, since in this case the information consists only in the position of the hit strip. 

In the assumption that two strips collect the whole particle signal, one can write down the following relationship ({\it linear $\eta$ algorithm}), which exploits the definition given in eq.~\ref{eq:etadef}, and corresponds to eq.~\ref{eq:ipr_cog} with the sum limited to two terms only:

\[ x = x_{L} + \eta \cdot P \]

\noindent
Here $x_{L}$ is the position of the left strip in the pair and $P$ is the \mbox{read--out} pitch. 

\item Unfortunately, due to the non--uniformity of the distribution of $\eta$, it can be proved that the position reconstructed by the linear $\eta$ algorithm is systematically shifted with respect to the true one~\cite{peisert:1992,landi:2002}. The systematically correct particle position is obtained by a {\it non--linear $\eta$ algorithm} containing a strictly increasing function $f(\eta)$:

\begin{equation} 
x = x_{L} + f(\eta) \cdot P 
\label{eq:ipr_etanonlin}
\end{equation} 

\noindent
The $f(\eta)$ function which produces the systematically correct impact point is the cumulative probability distribution function of $\eta$; it can be estimated as:

\begin{equation}
 f(\eta) = 
\ds{\int_{0}^{\,\eta} (dN/d \eta') \; d \eta'}\Bigg/ 
\ds{\int_{0}^{\,1} (dN/d \eta') \; d \eta'}
\label{eq:etaintdef} 
\end{equation}

\noindent
$dN/d \eta$ being the experimental $\eta$ distribution (fig.~\ref{fig:eta}), obtained by a  uniform lightening of the sensor~\cite{turchetta:1993}. This method, known in literature simply as {\it the $\eta$ algorithm}, has been extensively employed in the PAMELA data analysis: it allows to reach spatial resolutions of $3 \, \mu$m (junction side) and $12 \, \mu$m (ohmic side) for orthogonal tracks~\cite{nima478}. In our analysis the $\eta$ algorithm is applied if at least one of the channels adjacent to the maximum has a positive signal above the pedestal level; otherwise, the digital algorithm is used\footnote{Please be careful not to mistake the multiplicity of the cluster, defined in sect.~\ref{sec:reduction}, for the presence of positive channels around its maximum, requested to apply the algorithm.}.

\begin{figure}
\hspace*{-0.2cm}
 \begin{minipage}{7.5cm}
  \begin{center}
   \includegraphics[width=7cm, bb= 0 0 1534 2230, clip]{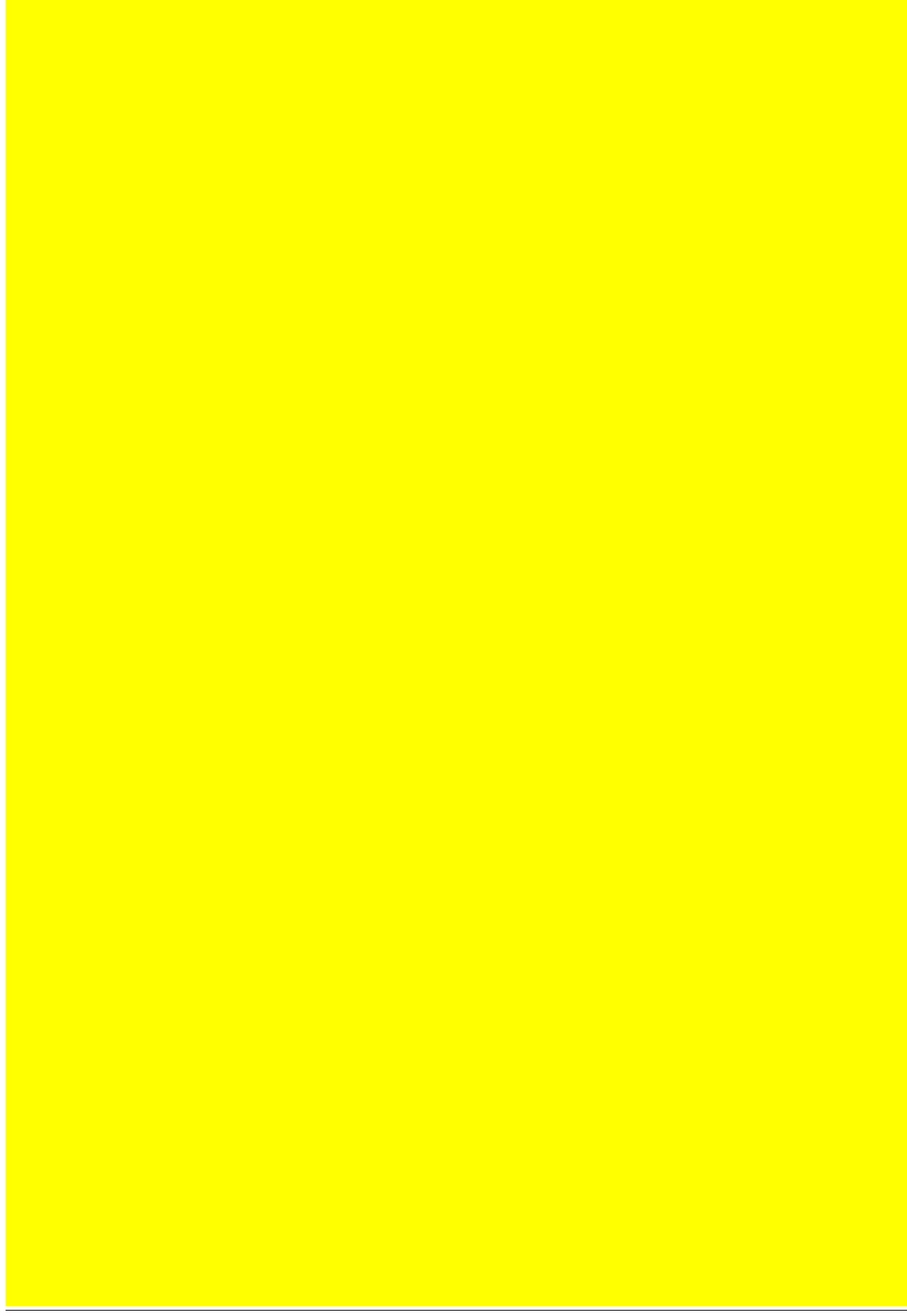}
  \end{center}
 \end{minipage}
\hspace*{-1.0cm}
 \begin{minipage}{7.5cm}
  \begin{center}
   \includegraphics[width=7cm, bb= 0 0 1534 2230, clip]{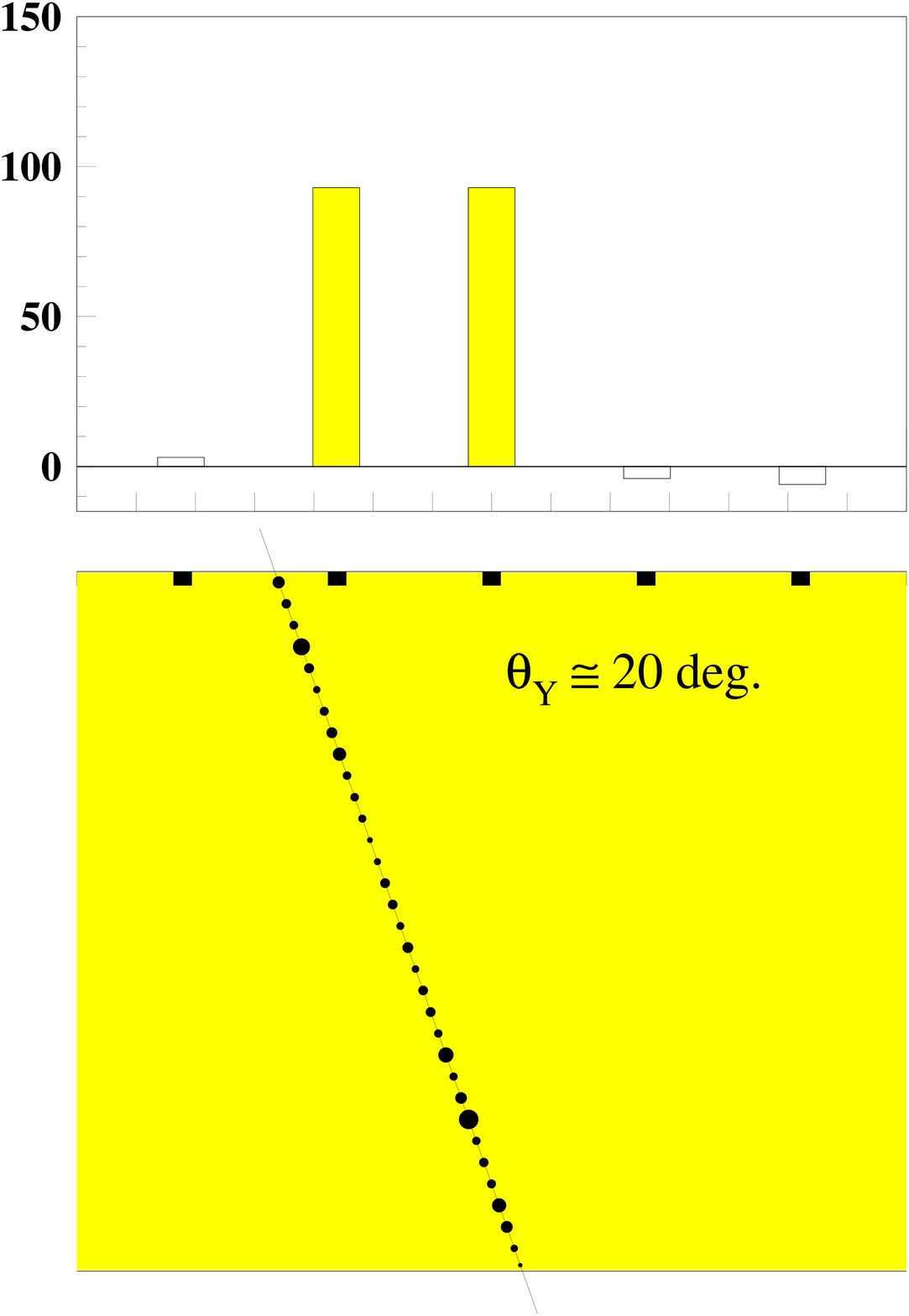}
  \end{center}
 \end{minipage}
 \begin{center}
\hspace*{+0.5cm}
  \begin{minipage}{12.5cm}
   \caption
{Simulation of an inclined track inside the silicon detector and corresponding clusters for junction ({\em left}) and ohmic ({\em right}) sides. Due to the different strip geometries, similar incidence angles result in different cluster multiplicities.} 
   \label{fig:incl_cl}
  \end{minipage}
 \end{center}
\end{figure}

\item In case of inclined tracks, the charge created in silicon can be collected by more than one or two strips, depending on the detector configuration (fig.~\ref{fig:incl_cl}). For some strip geometries (like that in the left picture in figure) it is natural to expect an improvement to the spatial resolution by an algorithm involving more than two strips in the determination of the impact point. For this reason, extensions of the non--linear $\eta$ algorithm can be thought, based on a larger number of strips. In particular, if three strips around the maximum are considered, a new variable can be introduced by exploiting the respective signals $S_{0}, \, S_{1}, \, S_{2}$:

\[ \eta_{3} = \frac{S_{1} + 2 \,S_{2}}{S_{0}+S_{1}+S_{2}} \]

\noindent
Similarly to what observed for $\eta$ (sect.~\ref{sec:tuning}), this new variable represents the weighted average of the positions of three strips, located in $0$, $1$, $2$; they are chosen so that the central one contains the maximum signal of the cluster. A further variable, based on the signals collected in four channels, can be conceived in a similar way:

\[ \eta_{4} = \frac{S_{1} + 2 \,S_{2} + 3 \,S_{3}}{S_{0}+S_{1}+S_{2}+S_{3}} \]

\noindent
In this case the strips are chosen so that $S_{1}, \,S_{2}$ are the biggest adjacent signals (those used in the $\eta$ algorithm) and $S_{0}, \,S_{3}$ are signals belonging to additional lateral strips (respectively on the left and right side of $S_{1}$ and $S_{2}$). According to their definitions, $\eta_{3}$ varies in the range $[0.5;1.5]$, while $\eta_{4}$ is contained in the interval $[1;2]$. A non--linear procedure, similar to that used for $\eta$, can be applied to improve the characteristics of this position finding algorithm extended to three or four strips: two functions $f_{3}(\eta_{3})$ and $f_{4}(\eta_{4})$ can be defined on the basis of the experimental distributions $dN/d\eta_{3}$ and $dN/d\eta_{4}$ as already done in eq.~\ref{eq:etaintdef}. The particle impact point will be consequently reconstructed as:

\begin{eqnarray*} 
x &=& x_{L} + f_{3}(\eta_{3}) \cdot P \hspace{1cm} \mbox{by the} \; \eta_{3} \; \mbox{algorithm} \\
x &=& x_{L} + f_{4}(\eta_{4}) \cdot P \hspace{1cm} \mbox{by the} \; \eta_{4} \; \mbox{algorithm}
\end{eqnarray*} 
  
\noindent
$x_{L}$ being the left strip in each group of three or four adjacent strips. \end{itemize}

Methods of impact point reconstruction for large incidence angles are also described in literature~\cite{turchetta:1993}; they are not treated here because of the limited angular acceptance of the PAMELA telescope, which corresponds to about $[-20;20]$~deg. for both coordinates. For such angles we expect in our strip geometry a cluster multiplicity of about three on the junction side and two on the ohmic side (fig.~\ref{fig:incl_cl}); fluctuations in these numbers, for similar incidence angles, can be produced by different positions of the track with respect to the strips. 

The methods presented in detail in the previous summary will be applied to the analysis of simulated events in the next sections.

\section{Position reconstructed by the $\eta$ algorithm: comparison with data} 
\label{sec:5planes}

\begin{figure}
 \begin{center}
  \begin{minipage}{7.5cm}
  \begin{center}
  \includegraphics[width=7cm]{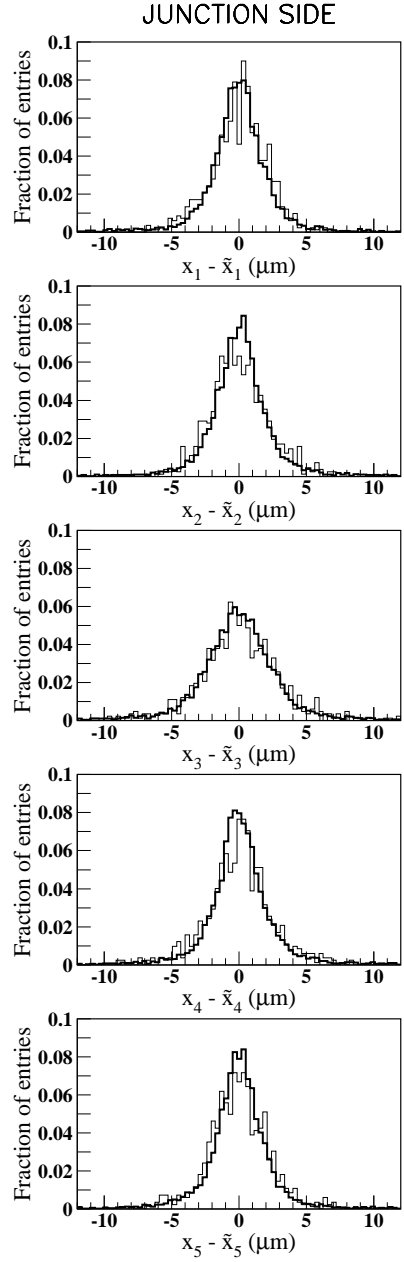}
  \caption{Distribution of the differences \mbox{$(x_{i}-\tilde{x_{i}})$} calculated for data (thin line) and simulation (thick line) with orthogonal tracks in absence of magnetic field. The experimental set--up consisted of five  non--uniformly spaced planes, arranged in a prototype telescope that acquired data on a test beam.}
  \label{fig:res5p}
  \end{center}
  \end{minipage}
 \end{center}
\end{figure}

Before applying the position finding algorithms, described in the previous section, to the analysis of simulated events, a comparison with data concerning the position reconstructed by the $\eta$ algorithm is suitable. $100 \; \mbox{GeV}$ electrons, with incoming direction orthogonal to the sensors, have been acquired without magnet on a test beam by a prototype telescope: it was composed of five planes, located, along the beam direction, in the origin, in $\pm 145$~mm and in $\pm 165$~mm. Among the gathered events, only those containing one cluster per view have been selected. In this case for each crossing particle five points $(x_{i},y_{i})$ can be found by the $\eta$ algorithm, corresponding to the best estimations of the impact coordinates on every plane. A linear fit on these points identifies five other points $(\tilde{x}_{i},\tilde{y}_{i})$ on the planes; the differences \mbox{$(x_{i}-\tilde{x_{i}})$} between reconstructed and interpolated positions on the junction side are reported for every detector in the histograms of fig.~\ref{fig:res5p} (thin line). In the simulation the experimental configuration has been reproduced, including the contribution from the multiple scattering on every silicon plane. The same differences with respect to a fitted straight line are reported in the histograms (thick line), as done for data. The good agreement between data and simulation is clearly visible. 

Let us notice that the RMS's of the distributions of two different planes are not comparable, since the detectors are not equidistant. Assuming the same uncertainty $\sigma$ for each reconstructed coordinate $x_{i}$, we can derive (as error propagation) the RMS of the distributions of \mbox{$(x_{i}-\tilde{x_{i}})$}. By a reverse procedure, from the five RMS values of fig.~\ref{fig:res5p} the spatial resolution of the detectors, defined as the  uncertainty on the reconstructed coordinates, can be obtained: from this data sample $\sigma = 3.0\, \mu$m~\cite{nima478} can be found.

\section{Spatial resolution for orthogonal and inclined tracks} 

The aim of this section is the study of the optimal position finding algorithm for our detectors, as obtained from the simulation of tracks incident at various angles. For this purpose minimum ionizing particles have been generated with incidence angles uniformly distributed in the $[-20;20]$~deg. range. This interval has been subdivided in $1$~deg. wide bins and consequently the simulated events have been classified on the basis of their incidence angles. The $\eta$ algorithm has been initially used on the whole angular range; a different $f(\eta)$ function (eq.~\ref{eq:etaintdef}) has been built for each angular bin starting from the $\eta$ distribution of the corresponding set and the particle position for every event has been finally reconstructed by exploiting eq.~\ref{eq:ipr_etanonlin}.

\begin{figure}
 \begin{center}
  \begin{minipage}{13.5cm}
  \begin{center}
  \includegraphics[width=14cm]{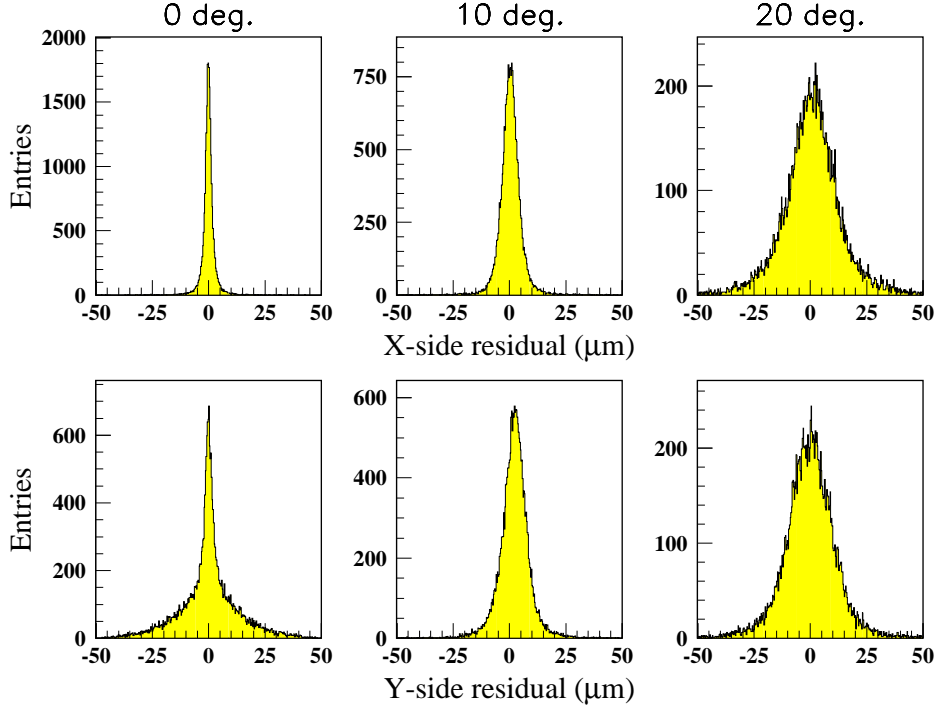}
  \caption{Distributions of residuals for the $\eta$ algorithm on junction ({\em top}) and ohmic ({\em bottom}) sides. The incidence angles of the particles are, from the left column to the right one, 0,~10,~20~deg.}
  \label{fig:res01020}
  \end{center}
  \end{minipage}
 \end{center}
\end{figure}
 
\begin{figure}
 \begin{center}
  \begin{minipage}{11cm}
  \begin{center}
  \includegraphics[width=8.5cm]{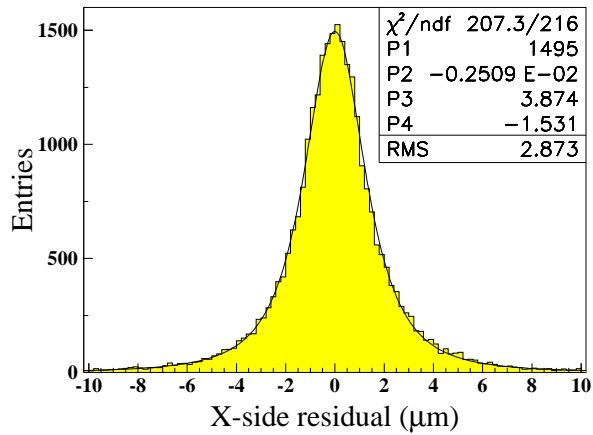}
  \caption{Simulated distribution of residuals on the junction side for orthogonal tracks. Parameters $p_{1}, \ldots, p_{4}$ are obtained from a fit with a generalized Lorentz distribution (eq.~\ref{eq:lorentz}).}
  \label{fig:xres}
  \end{center}
  \end{minipage}
 \end{center}
\end{figure}

In order to compare different algorithms, the distributions of the spatial residuals have been used: the residual is defined as the difference between the real particle impact point at half detector thickness, given by the simulation, and the reconstructed position, obtained by the chosen algorithm. For a quantitative comparison of different methods the RMS's of the distributions are evaluated. In fig.~\ref{fig:res01020} the distributions of the residuals, obtained from the $\eta$ algorithm on both junction and ohmic sides for $0, \, 10, \, 20$~deg. incidence angles, are shown. By simply comparing the distribution widths, it can be noticed that at small angles the junction side exhibits a better resolution, as expected, but at $20$~deg. the ohmic--side distribution is narrower: this is an indication of a loss of signal on the junction side out of the pair of strips the $\eta$ algorithm uses, since the multiplicity of the events becomes on the average greater than two when the track inclination increases (in tab.~\ref{tab:perc} the percentages of events with given multiplicities $m$ are reported for some incidence angles). We can expect that algorithms such as $\eta_{3}$ or $\eta_{4}$ will improve the resolution in this case. The non--Gaussian behaviour of the distributions (especially for small angles) can be observed too in fig.~\ref{fig:res01020}. On the junction side an accurate fit is provided for tracks perpendicular to the sensor by a generalized Lorentz distribution $L$ (fig.~\ref{fig:xres}):

\begin{equation}
 L = p_{1} \cdot \left[ 1 + \left( \frac{2 \cdot
(x-p_{2})}{p_{3}}\right)^{2\,} \right]^{p_{4}}
\label{eq:lorentz}  
\end{equation}

\noindent
The RMS of the distribution of residuals corresponds to the uncertainty in the reconstructed coordinates and consequentely can be defined as the spatial resolution of the detectors. The RMS of the distribution of fig.~\ref{fig:xres} ($2.87 \, \mu$m) can be compared with the spatial resolution ($3.0\, \mu$m) obtained for the real detectors from the RMS's of the distributions reported in fig.~\ref{fig:res5p}, as explained in sect.~\ref{sec:5planes}.

\begin{table}
 \begin{center}
\begin{tabular}{|ccc|c|c|c|c|c|c|c|}
\hline
\multicolumn{3}{|l|}{Incidence} & \multicolumn{4}{|c|}{junction side} & \multicolumn{3}{|c|}{ohmic side} \\
\cline{4-10}
\multicolumn{3}{|l|}{angle (deg.)} & $m = 1$ & $m = 2$ & $m = 3$  & $m \ge 4$ & $m = 1$ & $m = 2$ & $m \ge 3$ \\ 
\hline
\hline
$0$  & $\pm$ & $0.5$ & $35.4$~\% & $62.7$~\% & $ 1.5$~\% & $0.4$~\% & $81.6$~\% & $17.9$~\% & $0.5$~\% \\ 
$5$  & $\pm$ & $0.5$ & $22.9$~\% & $75.1$~\% & $ 1.6$~\% & $0.4$~\% & $75.5$~\% & $23.9$~\% & $0.6$~\% \\
$10$ & $\pm$ & $0.5$ & $ 6.6$~\% & $84.0$~\% & $ 8.8$~\% & $0.6$~\% & $61.4$~\% & $37.8$~\% & $0.8$~\% \\ 
$15$ & $\pm$ & $0.5$ & $ 1.3$~\% & $65.4$~\% & $32.3$~\% & $1.0$~\% & $43.4$~\% & $55.4$~\% & $1.2$~\% \\
$20$ & $\pm$ & $0.5$ & $ 0.4$~\% & $30.7$~\% & $66.4$~\% & $2.5$~\% & $24.2$~\% & $72.2$~\% & $3.6$~\% \\
\hline
\end{tabular}
\end{center}
\vspace*{0.3cm}
\caption{Percentage of events with given multiplicity $m$ as a function of the incidence angle.}
\label{tab:perc}
\end{table}

\noindent
On the ohmic side, where the pitch is larger, the evident tails in fig.~\ref{fig:res01020} in case of orthogonal tracks have been observed to be populated by some of the events in which the particle crosses the silicon detector some micrometers away from a strip and that collects all the released charge (in the Y--side  geometry this may happen even if the distance of the particle impact point from a strip is $10-15 \, \mu$m). In this case the left strip and the right one with respect to the maximum, that do not collect charge, can be included in the $\eta$ calculation (that always requires two strips) with equal probability, due to noise fluctuations in these channels. If the strip away from the real impact point is randomly chosen between the channels adjacent to the maximum, large errors in the reconstructed position will be observed. It can be easily verified that for such events the digital algorithm is comparable to $\eta$~\cite{samueletesi}.    

\begin{figure}
 \begin{center}
  \begin{minipage}{10cm}
  \begin{center}
  \includegraphics[width=7cm,bb=0 0 1534 2230, clip]{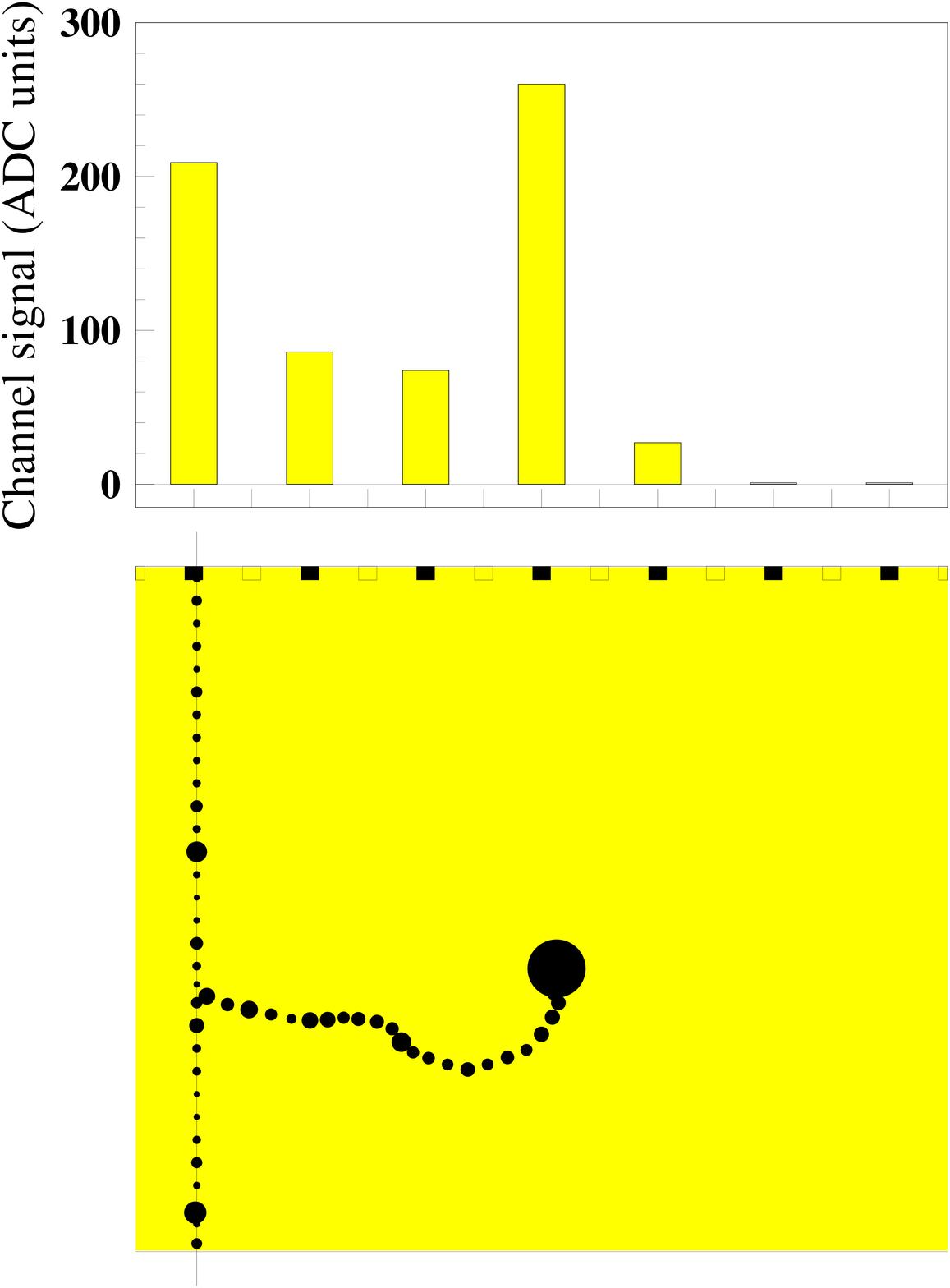}
  \caption{Track by a MIP with emission of a very energetic (about 260~keV) $\delta$ ray and corresponding cluster on the junction side, which exhibits two distinct peaks. The maximum signal corresponds to the Bragg's peak of the secondary electron.}
  \label{fig:cl_bad}
  \end{center}
  \end{minipage}
 \end{center}
\end{figure}

An independent contribution to the tails of the distributions on both sides come from the emission of $\delta$ rays in the detector (like that already shown in fig.~\ref{fig:evplot_o}). Another example of such an occurrence is shown in fig.~\ref{fig:cl_bad} along with the corresponding cluster signals: in this case the $\delta$ ray is so energetic that two different peaks appear in the cluster, the highest one being produced by the Bragg's peak of the secondary electron. The resulting error in the reconstructed position would be about $150\, \mu$m for this event, if the $\eta$ algorithm were applied around the maximum. In spite of this large error, such a cluster can be recovered in the final data analysis of the spectrometer, if two possible impact points are attributed to it, each corresponding to a cluster peak: the real position can be decided by comparison with the impact points on the remaining detecting planes during the tracking phase.

The behaviour of the spatial resolution as a function of the particle's incidence angle $\vartheta$ is now studied for a sample of tracks from which pathological events (like those above described, with emission of energetic $\delta$ rays) are excluded. Moreover a selection in the cluster multiplicity, depending on the track inclination, is introduced. The selection criteria are quite arbitrary, but they are employed only to produce a homogeneous sample of events where different algorithms are tested and compared. On the basis of the information contained in tab.~\ref{tab:perc}, the following clusters have been included in our analysis:

\begin{tabular}{ll}
\multicolumn{2}{l}{~~~~~~~~Junction side (bending view)} \\
$m=1,2$   & if $\vartheta \le 15$~deg. \\
$m=1,2,3$ & otherwise \\
\end{tabular}\\
\begin{tabular}{ll}
\multicolumn{2}{l}{~~~~~~~~Ohmic  side} \\
$m=1,2$  & for every $\vartheta$ \\
\end{tabular}

The $\eta$, $\eta_{3}$, $\eta_{4}$ and COG algorithms have been applied in every bin of incidence angle on the junction side; on the ohmic side only $\eta$ has been utilized, because of the small fraction of events whose multiplicity is greater than two. 

Let us now specify an aspect of the procedure that can give rise to some misunderstanding. When the $\eta$ algorithm is applied to clusters whose multiplicity is one, the charge released in silicon may be really collected by only one strip (the maximum of the cluster) and both the signals of the adjacent channels may be negative for noise fluctuations. In this case only the signal of the maximum is used in the impact point reconstruction and $\eta$ reduces to the digital algorithm. No restrictions are therefore introduced to such clusters, whose information is fully contained in the position of the maximum. The same criteria are followed for the other methods: as a general rule, when a sufficient number of strips with positive signals are not present in a cluster to apply a given algorithm, a procedure involving less channels is used instead, according to the sequence:

\begin{center}
\begin{tabular}{ccccccc} 
$\eta_{4}$ & $\rightarrow$ & $\eta_{3}$ & $\rightarrow$ & $\eta$ & $\rightarrow$ & digital algorithm.\\
\end{tabular} 
\end{center}

\begin{figure}
 \begin{center}
  \begin{minipage}{13.5cm}
  \begin{center}
  \includegraphics[width=14cm, bb=0 0 525 450, clip]{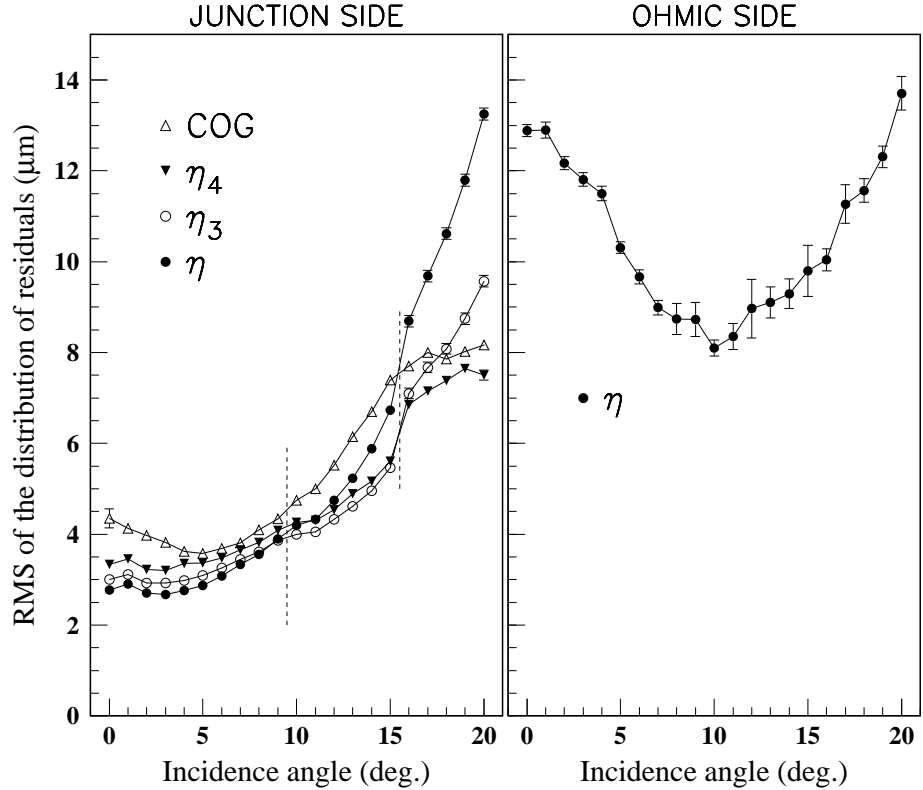}
  \caption{Spatial resolution obtained by some position finding algorithms as a function of the incidence angle on both junction and ohmic sides.}
  \label{fig:rms}
  \end{center}
  \end{minipage}
 \end{center}
\end{figure}

In fig.~\ref{fig:rms} the results provided by the mentioned methods on both  sides are shown. On the junction side it can be noticed that the $\eta$ algorithm gives the best resolution for orthogonal or slightly inclined tracks, as already stated in literature~\cite{turchetta:1993,peisert:1992,belau:1983}: in our configuration $\eta$ is the preferable method up to about $10$~deg. For larger angles, the original methods $\eta_{3}$ and $\eta_{4}$ give a more accurate particle position, compared to the Centre Of Gravity, suggested in the ref.~\cite{turchetta:1993} for intermediate incidence angles. In particular, $\eta_{3}$ can be used from about $10$ to $15$~deg. and $\eta_{4}$ from $15$ to $20$~deg. The spatial resolution ranges between less than $3 \, \mu$m and about $7.5\, \mu$m on the angular interval corresponding to the acceptance of the spectrometer. On the ohmic side a better resolution (about $8\, \mu$m RMS) is achieved for inclined ($\vartheta \sim 10$~deg.) tracks. This is due to a greater percentage of clusters with multiplicity $m = 2$ compared to case of perpendicular tracks, which result in a higher accuracy of the reconstructed position.

\section{The next step of this work}

Let us now briefly discuss how the spatial resolution, achieved at a given angle, influences the scientific objectives of the PAMELA experiment. The silicon microstrip detectors are the sensitive elements of a magnetic spectrometer that determines the momentum of the crossing charged particles. The obtained uncertainty $\Delta p/p$ in the reconstructed momentum is directly related, besides the magnetic field value and the geometry of the spectrometer, to the spatial resolution on the X (bending) view. Usually the performances of the instrument are given in terms of the rigidity $r$, defined as momentum--to--charge ratio: when $\Delta r/r = 100$~\% the spectrometer has reached the Maximum Detectable Rigidity (MDR). In PAMELA the value $\mbox{MDR} = ( 1183 \pm 54 ) \, \mbox{GV/}c$~\cite{boezio:2004} has been measured on a test beam for electrons whose arrival directions are orthogonal to the sensors. This value is influenced by the presence of tails in the distribution of residuals that can in principle be associated with errors in the charge sign of the reconstructed events. This problem becomes relevant in PAMELA because few antiparticles have to be identified in a large particle background (the antiparticle/particle ratio is of the order of $10^{-4}$). This drawback is known as {\it spillover} and actually limits the measurements of antiparticle abundances in PAMELA to about $190 \, \mbox{GeV/}c$ for antiprotons and $270 \, \mbox{GeV/}c$ for positrons~\cite{nima478}, in spite of a much larger MDR. For these reasons the data analysis needs accurate procedures that do not disperse the peculiar characteristics of the detectors. Results from this simulation will be therefore exploited to obtain the best algorithm of position reconstruction depending on the particle incidence angle and on the cluster multiplicity. Moreover, the knowledge of the width of the distribution of residuals in different conditions enables us to ``weight'' each impact point determination while tracking a particle through the spectrometer. For instance, a high--multiplicity cluster, occurring in a quasi--orthogonal track, is probably connected with a $\delta$ ray emission: so the corresponding position should reasonably have a lower weight in the track fit. If such a correlation is considered, the accuracy in the momentum reconstruction will be significantly improved.

\section{Conclusions} 

The simulation described in this paper takes into account the characteristics of the double--sided silicon microstrip detectors of the PAMELA spectrometer and correctly reproduces their observed performances. In particular the simulation has been applied to the analysis of the best obtainable spatial resolution as a function of the incidence angle of the incoming particle. New position finding algorithms ($\eta_{3}$ and $\eta_{4}$) have been successfully introduced to improve the resolution for non--orthogonal tracks on the bending view. The obtained distributions of the spatial residuals at different angles and multiplicities can be used in a simulation of the whole tracking system to study the particle contamination in antiparticle measurements and to calibrate in the best way the tracking procedure.


\end{document}